\documentclass[11pt,tightenlines,eqsecnum,floats,aps,prd,showpacs,amsmath,superscriptaddress,amssymb,nofootinbib,longbibliography]{revtex4-1}

\usepackage{bm}
\usepackage[latin1]{inputenc}
\usepackage[spanish,english]{babel}
\usepackage{amsfonts}
\usepackage{amssymb}
\usepackage{graphicx}
\usepackage{color}
\usepackage{stmaryrd}
\usepackage{mathtools,slashed}

\usepackage{amsmath}
\usepackage{soul}
\usepackage[normalem]{ulem}

\newcommand{\be}{\begin{equation}}
\newcommand{\ee}{\end{equation}}
\newcommand{\bea}{\begin{eqnarray}}
\newcommand{\eea}{\end{eqnarray}}
\newcommand{\ben}{\begin{enumerate}}
\newcommand{\een}{\end{enumerate}}

\begin{document}

\title{{\bf On the Electric-Magnetic Duality Symmetry: Quantum Anomaly, Optical Helicity, and Particle Creation\footnote{Review article prepared for  the Special Issue: ``Symmetry in Electromagnetism'', to appear in {\it Symmetry}\ (2018) 
}}}

\author{Ivan Agullo}\email{agullo@lsu.edu}\affiliation{Department of Physics and Astronomy, Louisiana State University, Baton Rouge, LA 70803-4001;}
\author{Adrian del Rio}\email{adriandelrio@tecnico.ulisboa.pt}
\affiliation{Centro de Astrof\'isica e Gravita\c{c}\~ao - CENTRA, Departamento de F\'isica, Instituto Superior T\'ecnico - IST, Universidade de Lisboa - 1049 Lisboa, Portugal;}
\author{Jose Navarro-Salas}\email{jnavarro@ific.uv.es}
\affiliation{Departamento de Fisica Teorica, IFIC. Centro Mixto Universitat de Valencia - CSIC.  Valencia 46100, Spain.}

\date{\today}

\begin{abstract}
It is well known that not every  symmetry of a classical field theory is also a symmetry of its quantum version. When this occurs, we speak of  quantum anomalies. The existence of anomalies  imply that  some classical Noether charges are no longer conserved in the quantum theory. In this paper, we  discuss a new example  for 
  quantum electromagnetic fields propagating in the presence of gravity. We argue that the symmetry under electric-magnetic duality rotations of the source-free Maxwell action is anomalous  in curved spacetimes. The classical Noether charge associated with these  transformations accounts for the net circular polarization or  the optical helicity of the electromagnetic field. 
Therefore, our results describe the way the spacetime curvature  changes the helicity of photons and opens the possibility of extracting   information from strong gravitational fields  through the observation of the polarization of photons. 
	We also argue that  the physical consequences  of this anomaly  can  be understood in terms of  the asymmetric quantum creation of photons by the gravitational~field. 


\end{abstract}


\maketitle

\section{Introduction}

Symmetries are at the core of well-established physical theories, and they keep playing a central role in the mainstream of current research. Fundamental Lagrangians in physics are  founded on symmetry principles. Moreover, symmetries are linked, via Noether's theorem, to conservations laws. Well-known examples are the  energy and momentum conservation and its relation with  the invariance under space--time translations, as well as the conservation of the net fermion number (the difference in the number of fermions and anti-fermions  that is proportional to the net electric charge) in   Dirac's relativistic  theory, which 
 result from the global phase invariance of the action. 

When the symmetries of free theories are also preserved by interactions, the conservation laws are maintained, and they can be used to understand patterns in diverse physical phenomena. In quantum electrodynamics, for instance, the phase invariance is preserved by the coupling of the Dirac and the electromagnetic field, and this ensures the conservation of the net fermion number in all physical processes \cite{qftbook}. Another illustrative example is the gravitationally induced creation of particles, either bosons or fermions, in an expanding homogenous universe \cite{parkerthesis, parker68, parker69, parker71}. This particle creation occurs in pairs, and the symmetry of the  background under space-like translations ensures that, if one particle is created with wavenumber  $\vec k$, its partner has wavenumber $-\vec k$. As a consequence, there is no creation of net momentum, as expected on symmetry grounds. In a similar way, phase invariance implies that the gravitational field cannot create a net fermion number in an expanding universe.

However, in special cases, the implications of classical symmetries do not extend to   quantum theory, and the classical charge conservation breaks down.
This was first noticed by studying  massless fermions coupled to an electromagnetic field \cite{ABJ, BJ}.  A massless fermion is called a (Weyl) left-handed fermion if it has helicity $h=-1/2$, and a right-handed fermion if $h=+1/2$~(A left-handed (right-handed) anti-fermion has helicity $h=+1/2$ ($-1/2$)). Recall that the equations of motion for the two {sectors}  decouple in the massless limit, and this allows one to  write a theory for massless fermions that involves only one of the two helicities, something that is not possible for non-zero mass. \mbox{The action} of this theory also enjoys   phase invariance, so the number of left-handed  and right-handed fermions is {\em separately} conserved. This is to say, in the classical theory, there are two {independent} Noether currents, $j^\mu_L$ and $j^\mu_R$, associated with left- and right-handed massless fermions, respectively,  that satisfy continuity equations $\partial_\mu j^\mu_L=0$ and $\partial_\mu j^\mu_R=0$. Rather than using $j^\mu_L$ and $j^\mu_R$, it is more common to re-write these conservation laws in terms of the so-called vector and axial currents, defined by their sum and difference, respectively, $j^\mu = j^\mu_R+ j^\mu_L= \bar \psi \gamma^\mu\psi$ and  $j^{\mu}_{5} = j^\mu_R - j^\mu_L= \bar \psi \gamma^\mu\gamma^5\psi$, where  $\psi$ is the four-component Dirac spinor, that encapsulates both left- and right-handed (Weyl) fermions, and $\gamma^{\mu}$, $\gamma_5$ are the  Dirac matrices.

What is the situation in   quantum theory? It turns out that the conservation law for $j^\mu$ holds also quantum mechanically, so the quantum number $N_R + N_L$ (associated with the net fermion number, i.e., the electric charge)  is preserved in any physical process. For instance, in the presence of a time-dependent electromagnetic background,  charged fermions and antifermions  are spontaneously created (this is the electromagnetic analog of the gravitationally induced particle creation mentioned above \cite{Schwinger, dunne, ferreiro}),  
but in such a way that the total fermion number (or electric charge) does not change. This is because the number of created right- or left-handed  antifermions equals the number of left- or right-handed fermions:
\bea N_R+N_L &=&  (\#_{1/2}^R - \#_{-1/2}^R )  + (\#^L_{-1/2} - \#^L_{1/2} ) = 0 \ . \eea

However, it turns out that the difference in the created number of right-handed and left-handed fermions is not identically zero. This means that it is possible to create a net amount of helicity:
\bea N_R-N_L &=&  (\#_{1/2}^R - \#_{-1/2}^R )  - (\#^L_{-1/2} - \#^L_{1/2} ) =  [\#_{1/2}- \#_{-1/2} ] \ . \eea

The simplest scenario where this is possible is for a constant  magnetic field, say, in the third spatial direction $\vec B= (0, 0, B)$, together with a pulse of electric field parallel to  it, $\vec E  = (0, 0, E(t))$. One can show that, in this situation, the net creation of helicity per unit volume $V$    is given by (see \cite{qftbook} for a proof involving an adiabatic electric pulse)
\be \label{DeltaNRNL}\frac {\Delta(N_R-N_L)}{V}= -\frac{q^2}{2\pi^2} \int_{t_1}^{t_{2}} dt \ \vec E \cdot \vec B \ , \ee
where $q$ is the electric charge of the fermion, and the fermionic field is assumed to start in the vacuum state at early times before the electric field is switched on.  Hence, if the integral $ \int_{t_1}^{t_{2}} dt \ \vec E \cdot \vec B $ is {different}  from zero, particles with different helicities are created in  different amounts. 
In contrast, $N_R+N_L$ remains strictly constant. 
For an arbitrary electromagnetic background, the previous result generalizes~to
\be \label{DeltaNRNL2} \Delta(N_R-N_L)= -\frac{q^2}{2\pi^2} \int_{t_1}^{t_{2}} dt \int d^3x \ \vec E \cdot  \vec B \ . \ee 
The key point is that Equation (\ref{DeltaNRNL2}) is equivalent to the quantum-mechanical symmetry breaking of the fermion chiral symmetry: $\psi \to \psi'= e^{-i \epsilon \gamma^5} \psi$, as  expressed in the anomalous non-conservation of the current  \cite{ABJ, BJ}. 
\be \label{ABJ} \partial_\mu \langle j^{\mu}_{5} \rangle = -\frac{q^2\hbar}{8\pi^2} F_{\mu\nu} \ {^{*}}F^{\mu\nu} \  \ee 
where $F_{\mu\nu}$ is the electromagnetic field strength,  $^{*}F_{\mu\nu}\equiv  \frac{1}{2} \epsilon_{\mu\nu\alpha\beta}F^{\alpha\beta}$ its dual, and the presence of $\hbar$ makes  manifest that this is a quantum effect. Equations (\ref{ABJ}) and (\ref{DeltaNRNL2}) are  connected by the standard relation between a current and the charge associated with it: $\int d^3x \langle j^{0}_{5} \rangle = \hbar  (N_R-N_L)$.

The discovery of the chiral anomaly of Equation (\ref{ABJ}) was not arrived at by computing the number of fermions created, but rather by directly computing the quantity $\partial_\mu \langle j^{\mu}_{5} \rangle$. In that calculation, the anomaly arises from the renormalization subtractions needed to calculate the expectation value   $\langle j^{\mu}_{5} \rangle$. The operator  $j^{\mu}_{5}$ is non-linear (quadratic)  in the fermion field, and in quantum field theory expectation values of non-linear operators are plagued with ultraviolet divergences. One must use renormalization techniques to extract the physical, finite result. A detailed study shows that renormalization methods that respect the gauge invariance of the electromagnetic background break the fermionic chiral symmetry of the classical theory. The fact that Expression (\ref{ABJ}) was able to accurately explain the decay ratio of processes that could not be understood otherwise, like 
 the decay of the neutral pion to two photons, was an important milestone in the quantum field theory and the study of anomalies.



A similar anomaly appears when the electromagnetic background is replaced by a gravitational field \cite{kimura}. In this case, renormalization methods that respect general covariance give rise to a violation of the classical conservation law $\nabla_\mu  j^{\mu}_{5} =0$, which becomes
\be \label{Kimura} \nabla_\mu \langle j^{\mu}_{5} \rangle = -\frac{\hbar}{192\pi^2} R_{\mu\nu\alpha\beta}\ {^{*}}R^{\mu\nu\alpha\beta} \  , \ee 
where $R_{\mu\nu\alpha\beta}$ is the Riemann tensor and ${^{*}}R^{\mu\nu\alpha\beta}$ its dual, and $\nabla_{\mu}$ is the covariant derivative. 
\mbox{For gravitational} fields for which a particle interpretation is available at early and late times, \mbox{this chiral}  anomaly also manifests in the net helicity contained in the fermionic  particles created during the~evolution:
\be \label{DeltaNRNL2gravity} \Delta(N_R-N_L)= -\frac{1}{192\pi^2} \int_{t_1}^{t_{2}} \int_{\Sigma} d^4x\sqrt{-g} \ R_{\mu\nu\alpha\beta}\ {^{*}}R^{\mu\nu\alpha\beta} \ . \ee

 Typical configurations where this integral is  non-zero are the gravitational collapse of a neutron star, or the merger of two compact objects as the ones recently observed by the LIGO-Virgo collaboration \cite{Abbot1, Abbot2}.   
 
In contrast to the anomaly of Equation (\ref{ABJ}) induced by an electromagnetic background, the chiral anomaly induced by gravity affects every sort of massless spin-$1/2$ fields, either charged or neutral. This is a consequence of the universal character of gravity, encoded in the  equivalence principle, that guarantees that, if Equation (\ref{Kimura}) is valid for a type of massless spin-$1/2$ field, it must also be valid for any other type.

There is no reason to believe, however, that these anomalies are specific  to spin $1/2$ fermions, \mbox{and one} could in principle expect that a similar effect will arise for other types of fields that classically  admit chiral-type symmetries. This is the case of photons. One then expects that photons propagating in the presence of a  gravitational field will not preserve their net helicity, or, in the language of particles, \mbox{that the} gravitational field will created photons with different helicities in unequal amounts,  in the same way   it happens for fermions~(one also expects a similar effect for gravitons). For photons, \mbox{the analog} of the classical chiral symmetry of fermions is given by  electric-magnetic duality rotations \cite{jacksonbook}
\bea \label{drotations}\vec E ' &=& \cos\theta \ \vec E + \sin\theta \ \vec B \nonumber \\
\vec B' &=& \cos\theta \ \vec B - \sin\theta \ \vec E  \ . \eea

As first proved in \cite{calkin1965invariance}, these transformations leave the action of electrodynamics invariant if sources (charges and currents) are not present, and the associated Noether charge is precisely the difference between the intensity of the right- and left-handed circularly polarized electromagnetic waves, i.e., the net helicity. This symmetry is exact in the classical theory even in the presence of an arbitrary gravitational background, as pointed out some years  later in \cite{deser1976duality}. 
 However, in exact analogy with the fermionic case, quantum effects can break this symmetry of the action and induce an anomaly \cite{essay, ADRNS17}. In the language  of particles, this would imply that  the difference in the number of photons with  helicities $h=\pm1$, $N_R-N_L$,  is not necessarily conserved in curved spacetimes. The analogy with the fermionic case also suggests that the current $j_D^{\mu}$ associated with the symmetry under duality rotations of the classical theory fails to be conserved quantum mechanically, with a non-conservation law of the type
%
\be \label{anomalyD} \nabla_\mu \langle j_D^{\mu} \rangle = \alpha\,  \hbar R_{\mu\nu\alpha\beta}\ {^{*}}R^{\mu\nu\alpha\beta} \ , \ee 
 where  $\alpha$ is a numerical coefficient to be determined. In a recent work \cite{ADRNS17, ADRNS18}, we have proved that this is in fact the case, and obtained that $\alpha$ is different from zero and 
  given by $\alpha= -\frac{1}{96\pi^2}$. In this paper, we will provide a general overview of these results from a different perspective, and with more emphasis on conceptual aspects.

\section{Electric-Magnetic Duality Rotations and Self- and Anti Self-Dual Fields} \label{Electrodynamics}

To study electric-magnetic rotations of Equation (\ref{drotations}), it is more convenient to change variables to the self- and anti-self-dual components of  the electromagnetic field, defined by $ \vec H_{\pm}\equiv\frac{1}{\sqrt{2}} \, (\vec E\pm i\, \vec B)$, since for them 
the transformation of Equation (\ref{drotations}) takes a diagonal form:
\be \vec H'_{\pm}   = e^{\mp \, i \theta} \vec H_{\pm}\, . \ee

A discrete duality transformation $\star \vec E=\vec B$, $\star \vec B=-\vec E$ corresponds to $\theta=\pi/2$. 
Then, the operator $i\star$ produces 
$i\star\, \vec H_{\pm} =\pm\, \vec H_{\pm}$. It is for this reason that $\vec H_+$ and $\vec H_-$ are called the self- and anti-self-dual components of the electromagnetic field, respectively.

There are other aspects that support the convenience of these variables. For instance, under a Lorentz transformation, the components of  $\vec E$ and $\vec B$ mix with each other. Indeed, under an infinitesimal Lorentz transformation of rapidity  $\vec \eta$ 
, the electric and magnetic fields transform as $\vec E' =\vec E -\vec \eta \wedge \vec B, \vec B' =\vec B +\vec \eta \wedge \vec E$. (We recall that the rapidity  $\vec \eta$ completely characterizes a Lorentz boost: its modulus contains the information of the  Lorentz factor  $\gamma$, via $\cosh{|\vec \eta|}=\gamma$, and its direction indicates  the direction of the boost). However, when  $\vec E$ and $\vec B$ are combined into $\vec H_{\pm}$, it is easy to see that the components of $\vec H_+$ and $\vec H_-$ no longer mix:
 \be\label {law} \vec H'_{\pm} = \vec H_{\pm} \ {\pm}\ i \vec \eta \wedge \vec H_{\pm} \ . \ee
 
 Note also that, under an ordinary infinitesimal (counterclockwise) rotation of angle $\alpha >0$ around the direction of a unit vector $\vec n$, the complex vectors $\vec H_{\pm}$ transform as $ \vec H'_{\pm} = \vec H_{\pm} + \alpha \vec n \wedge \vec H_{\pm}$.
 Hence, a boost corresponds to a rotation of an imaginary angle. These are the transformation  rules associated with the two irreducible representations of the Lorentz group for fields of spin $s=1$. In the standard terminology \cite{Weinbergqft, AGVM}, they correspond to the $(0,1)$ representation for  $\vec H_+$, and the $(1,0)$ one for $\vec H_-$. More generally, for any element of the  restricted Lorentz group $SO^+(1, 3)$ (rotations + boots), the above complex fields transform as
 \be\vec H'_{\pm}= e^{-i(\alpha \vec n \ {\pm}i\vec \eta ) \cdot \vec J} \, \vec H_{\pm} \   \ee
where $\vec J$ are the infinitesimal generators of the group of rotations.  The ${\pm}$ sign in the above equation distinguishes the two inequivalent (three-dimensional) representations of the Lorentz group. \mbox{They are,} however, equivalent under the  subgroup of rotations. This makes transparent the fact that electrodynamics describes fields of spin $s=1$, something that is more obscure when working with $\vec E$ and $\vec B$,  the field strength $F_{\mu\nu}$, or even the vector potential $A_{\mu}$. 

Another   useful aspect of self- and anti-self-dual variables concerns the equations of motion. \mbox{The source-free} Maxwell equations  
\bea \label{Meq} \vec \nabla \cdot \vec E&=&0\, , \hspace{2cm} \vec \nabla \cdot \vec B =0 \;  \nonumber \\
\vec \nabla \times \vec E &=&-\partial_t\, \vec B \, , \hspace{1cm} \vec \nabla \times \vec B =\partial_t\, \vec E\, 
\eea
when written in terms of $\vec H_{\pm}$, take the form
\be \label{HMeq} \vec \nabla \cdot \vec H_{\pm}=0\, , \hspace{1cm} \vec \nabla \times \vec H_{\pm} =\pm i\, \partial_t\, \vec H_{\pm} \, .
\ee

Notice that, in contrast to $\vec E$ and $\vec B$, the self- and anti-self-dual fields are not coupled by the dynamics. 
The general solution to these field equations is a linear combination of positive and negative frequency plane waves
\be \label{solH} \vec H_{\pm}(t,\vec x)=\int \frac{d^3 k}{(2\pi)^3}\, \Big[ h_{\pm}(\vec k)\, e^{-i (k\, t-\vec k\cdot \vec x)}+ h^*_{\mp}(\vec k)\, e^{i (k\, t-\vec k\cdot \vec x)} \Big] \, \hat \epsilon_{\pm}(\vec k)\, ,  \ee
where $k=|{\vec k}|$ and  $h_{\pm}(\vec k)$ are complex numbers that quantify the wave amplitudes. 
The polarization vectors are given by $\hat \epsilon_{\pm}(\vec k)=\frac{1}{\sqrt{2}}\, (\hat e_1(\vec k) \pm i\, \hat e_2(\vec k))$ where  $\hat e_1$ and $\hat e_2$ are any two real, space-like unit vectors  transverse to $\hat k$ (we choose their orientation such that  $\hat e_1 \times \hat e_2=+\hat k$).  Positive-frequency Fourier modes  $h_{\pm}(\vec k)\, e^{-i (k\, t-\vec k\cdot \vec x)}\, \hat \epsilon_{\pm}(\vec k)$ describe waves with helicity $h=1$ for self-dual fields, and with negative helicity  $h=-1$ for anti-self-dual fields.  
This  is also in agreement with the general fact that a massless field associated with the Lorentz representation $(0, j)$ describes particles with helicity $+j$, while a $(j,0)$-field  describes particles with helicity $-j$ \cite{Weinbergqft}. Compared with massless fermions,  $\vec H_+$ is the analog of a right-handed Weyl spinor, which transforms 
  under the $(0,1/2)$ Lorentz representation, 
  and $\vec H_-$ is the analog of a left-handed Weyl spinor.

The constraints $\vec \nabla \cdot \vec H_{\pm}=0$ can be used to introduce the potentials  $ \vec A_{\pm}$, as follows:
\be \label{defApm} \vec H_{\pm}=\pm \, i\, \vec \nabla \times \vec A_\pm  . \ee
Maxwell equations  then reduce to first-order differential equations for the potentials:
\be \label{AMeq} \pm i \, \vec \nabla \times \vec A_{\pm} = -\, \partial_t\, \vec A_{\pm} + \vec \nabla A^0_{\pm} \, .\ee
Both  sets of equations,  for the fields of Equation (\ref{HMeq})   and  for the  potentials of Equation (\ref{AMeq}), can be written more compactly as follows (the equations for $ \vec H_-$ and $\vec A_{-\, }$ are obtained by complex conjugation)%
\be \label{1oeq}  \alpha^{ab}_{I} \partial_{a} H^I_+=0\, , \hspace{1cm} \bar \alpha^{ab}_{I} \partial_{a} A_{+\, b}=0 \, . \ee
The numerical constants $\alpha^{ab}_{I} $ are three $4\times4$ matrices, for $I=1,2,3$, and the bar over $\alpha^{ab}_{I}$ indicates complex conjugation. The components of these matrices in an inertial frame are
 \bea
\alpha^{ab}_1 =  \left( {\begin{array}{cccc}
  0 &- 1 & 0& 0  \\
1 & 0 & 0 & 0  \\
0 & 0 & 0 & i  \\
0 & 0 & -i & 0  
  \end{array} } \right)  \, 
\hspace{0.3cm} 
\alpha^{ab}_2  = \left( {\begin{array}{cccc}
  0 & 0 & -1& 0  \\
0 & 0 & 0 & -i  \\
1 & 0 & 0 & 0 \\
0 & i & 0 & 0  
  \end{array} } \right) \, 
  \hspace{0.3cm}
  \alpha^{ab}_3  =   \left( {\begin{array}{cccc}
  0 & 0 & 0& -1  \\
0 & 0 & i & 0  \\
0 & -i & 0 & 0 \\
1 & 0 & 0 & 0  
  \end{array} } \right)
 \ .  \eea

  It is trivial to check by direct substitution that Equation  (\ref{1oeq}) is equivalent to Equations (\ref{HMeq})  and (\ref{AMeq}), respectively. 
 These anti-symmetric matrices are Lorentz invariant symbols. They are self-dual ($i \star \alpha^{ab}_I=\alpha^{ab}_I$), and the conjugate matrices are  anti-self-dual  ($i \star \bar \alpha^{ab}_I=- \bar \alpha^{ab}_I$). 
  The two sets of equations in Equation (\ref{1oeq}) were shown  in \cite{ADRNS18} to contain the same information. One can thus formulate source-free Maxwell theory entirely in terms of complex potentials.

\section{Noether Symmetry and Conserved Charge}\label{section2}

In this section, we show that electric-magnetic rotations of Equation (\ref{drotations}) are a  symmetry of the classical theory, and  obtain an expression for the associated conserved charge. This can be more easily done by working in   Hamiltonian formalism. The phase space of electrodynamics is usually parametrized by the pair of fields $(\vec A (\vec x), \vec E (\vec x))$, with $\vec B = \vec \nabla \times \vec A$. The Hamiltonian of the theory is easily obtained by the Legendre transform from 
the  Lagrangian, and it reads
 \be H= \frac{1}{2}\int d^3x \,  \Big[\vec{E}^2 + (\vec \nabla \times \vec A)^2-A_0\,  (\vec \nabla \cdot \vec E)\Big]\, .\ee
 In this expression, $A_0(\vec{x})$ is regarded as a Lagrangian multiplier that enforces the Gauss law constraint $\vec \nabla \cdot \vec E=0$.  The phase space is equipped with a Poisson structure given by $\{A_i(\vec{x}),  E^j(\vec{x}')\}=\delta_i^j \, \delta^{(3)}(\vec{x}-\vec{x'})$, which induces a natural symplectic product $$\Omega[(\vec A_1, \vec E_1),(\vec A_2, \vec E_2)]=-\frac{1}{2}\int d^3 x \left[ \vec E_1 \cdot \vec A_2- \vec E_2 \cdot \vec A_1\right] \ . $$  
From the form of the electric-magnetic rotations of Equation (\ref{drotations}), we see that the  infinitesimal transformation of the  canonical variables reads
\bea \label{dtps2}\delta \vec A= \vec Z\, , \hspace{1cm}   \delta \vec E=\vec \nabla \times \vec A\,   \eea
 where $\vec Z$ is defined by $\vec E =: -\vec \nabla \times \vec Z$; therefore, it can be understood as an ``electric potential'' (note that  in the source-free theory $\vec Z$ can be always defined, since $\vec \nabla \cdot \vec E=0$).
 
Now, the  generator of the transformation  of Equation (\ref{dtps2}) can be determined by  %
 \be \label{Cargadual2} Q_D=\Omega[(\vec A, \vec E),(\delta \vec A, \delta \vec E)]=- \frac{1}{2}\int d^3x\, [\vec E\cdot \delta \vec A-\vec A\cdot  \delta \vec E]=\frac{1}{2}\int d^3x\, [\vec A\cdot  \vec B- \vec Z \cdot \vec E]\, . \ee
 $Q_D$ is gauge invariant, and one can easily check that it generates the correct transformation by computing Poisson brackets
\bea \delta \vec B&=& \{\vec B,Q_D\} =   \{\nabla\times \vec A,Q_D\}=-\vec E \   \nonumber \\
\delta \vec E&=&    \{\vec E,Q_D\} =\vec B\ . \eea
%
It is also straightforward to check that $\delta H=\{ H, Q\}=0$.  Therefore, the canonical transformation generated by $Q_D$, i.e., the electric-magnetic duality transformation of Equation (\ref{dtps2}), is a symmetry of the source-free Maxwell theory, and $Q_D$ is a constant of motion. 

Taking into account the  form of the generic solutions,  Equation (\ref{solH}), to the field equations, the conserved charge reads
\be Q_D=\int \frac{d^3k}{(2\pi)^3\, k}  \Big[ |h_+(\vec k)|^2-|h_-(\vec k)|^2\Big] \, . \label{cargadual3}\ee
This expression makes it  clear that $Q_D$ is proportional to the difference in the intensity of the self- and anti-self-dual parts of field or, equivalently, the difference between the right and left circularly polarized components. In the quantum theory, $Q_D/\hbar $ measures the difference in the number of photons with helicities $h=+1$ and  $h=-1$.   {For this reason,   we recognize $Q_D$ as the V-Stokes parameter that describes  the polarization state of the electromagnetic radiation.  

Although we have restricted here to Minkowski spacetime, the argument generalizes to situations in which a gravitational field is present  \cite{deser1976duality}. A generally covariant proof in curved spacetimes in the Lagrangian formalism is given in  \cite{ADRNS18}, where the associated Noether current was obtained: 
\bea
j^{\mu}_D= \frac{ 1}{2}\,   \Big[  A_{\nu}\, {^\star F}^{\mu\nu}-Z_{\nu}\,  F^{\mu\nu}\Big] \, .
\eea

\section{   Analogy with Dirac Fermions and the Quantum Anomaly }\label{Section}

The goal of this section is to compute  the vacuum expectation value of the current $j^{\mu}_D$ associated with the symmetry under electric-magnetic rotations, and to use the result  to evaluate whether these transformations are also a symmetry of the quantum theory. A convenient strategy to achieve this  is to realize that, in the absence of electric charges and currents, Maxwell's theory can be formally written as a (bosonic)  spin $1$  version of the Dirac theory for a real spin $1/2$ field. The convenience of writing the theory in this form is that it  allows one to take advantage of numerous and powerful tools developed to compute the chiral anomaly for fermions. Hence, we will start in Section  \ref{s.4.1} by summarizing the theory of massless spin $1/2$ fermions and the calculation of the fermionic chiral anomaly, and   we will come back to the electromagnetic case in Section \ref{s.4.2}.

\subsection{Fermions in Curved Spacetime\label{s.4.1}}

To better motivate the analogy between  electric-magnetic rotations and  chiral rotations of fermions, it is convenient to write the Dirac field in terms of two Weyl spinors  $\psi^{}_L$ and $\psi^{}_R$ as follows (see for \mbox{instance \cite{qftbook, Srednikybook}}):
\bea
\psi  \equiv \left( {\begin{array}{c}
    \psi_L  \\
  \psi_R\\
  \end{array} } \right)\ , \ \ \ \ \ \ \  \bar\psi \equiv \psi^{\dagger} \beta = ( \psi^{\dagger}_L, \psi^{\dagger}_R ) \   \  \label{psi}
\eea 
where $\beta$ is the matrix
\be \beta\equiv \left( {\begin{array}{cc}
0 & I \\
I  &0
  \end{array} } \right) . \ee
  
The spinor $\psi_L$ transforms according to the $(1/2, 0)$ representation of the Lorentz algebra, while the spinor $\psi_R$ transforms with the $(0, 1/2)$ representation.  The Dirac equation 
\be i \gamma^{\mu} \partial_\mu \psi= m \psi \  \ee  takes the form
\bea
 i \left( {\begin{array}{cc}
 0 &  \sigma^{\mu}  \\
  \bar\sigma^{\mu} & 0 \\
  \end{array} } \right)     \left( {\begin{array}{c}
    \psi_L  \\
  \psi_R\\
  \end{array} } \right) =  m   \left( {\begin{array}{c}
    \psi_L  \\
  \psi_R\\
  \end{array} } \right) \  ,
\eea
where $\sigma^{\mu}= (I, \vec \sigma)$ and $\vec \sigma$ are the Pauli matrices. Numerically $\beta$ agrees with the Dirac matrix $\gamma^0$, and it is for this reason that the two matrices are commonly identified (although they have a different index structure; see e.g.,\ \cite{Srednikybook}).  For massless fermions, the theory is invariant under the chiral transformations $\psi \to \psi'=e^{i\theta \gamma_5} \psi$, with $\gamma_5= \frac{i}{4!}\epsilon_{\alpha\beta\gamma\delta}\gamma^{\alpha}\gamma^{\beta}\gamma^{\gamma}\gamma^{\delta}$ 
\be \gamma_5 =  \left( {\begin{array}{cc}
-I& 0  \\
0 & I  \end{array} } \right)   \ .   \ee
Therefore, 
\bea
\psi= \left( {\begin{array}{c}
    \psi_L \\
  \psi_R\\
  \end{array} } \right) \to    \psi' = e^{i\gamma_5 \theta} \psi = \left( {\begin{array}{c}
   e^{-i\theta} \psi_L \\
 e^{i\theta} \psi_R\\
  \end{array} } \right)
\  . \eea
Noether's theorem associates with this symmetry transformation the chiral current $j^\mu_5= \bar \psi \gamma^\mu \gamma_5\psi$. The spatial integral of its  time-component is the charge 
\be Q_5= \int d^3x (\psi_R^{\dagger}\psi_R - \psi_L^{\dagger}\psi_L) \ , \ee
and it is classically conserved.
This   charge counts the difference in the number of positive and negative helicity states,  in close analogy to the dual charge of Equation (\ref{cargadual3}) for the electromagnetic case. As we mentioned in the introduction,  this quantity is  a constant of motion in the quantum theory in Minkowski space, but this is not necessarily  true in the presence of a gravitational  background, as  we now explain  in more detail.

In  the presence of gravity,  the Dirac equation for a massless spin $1/2$ fields takes the form (see, for instance, \cite{Parker-Toms}) 
\be i \gamma^\mu (x) \nabla_\mu \psi (x) =0 \ , \ee
where  $\gamma^\mu (x) = e^\mu_a (x) \gamma^a$ are the Dirac gamma matrices in curved space,  $ e^\mu_a (x) $ is a Vierbein or orthonormal tetrad in terms of which the curved metric $g_{\mu\nu}$ is related to the Minkowski metric $\eta_{ab}$ by $g_{\mu\nu}\, e^\mu_a e^\nu_b= \eta_{ab}$, while $\gamma^a$ are the Minkowskian gamma matrices (that satisfy $\{\gamma^a, \gamma^b\}= 2\eta^{ab}$). $\nabla_\mu$ is the covariant derivate acting on spin $1/2$ fields:
\be \nabla_\mu \psi = (\partial_\mu + i\omega_{\mu ab}\Sigma^{ab}) \psi \ , \ee
where $\Sigma^{ab}= -\frac{1}{8} [\gamma^a, \gamma^b]$ are   the generators of the $(1/2, 0) \bigoplus (0, 1/2) $ representation of the Lorentz group, 
and $w_{\mu}$ is the standard spin connexion, defined in terms of the Vierbein and the Christoffel symbols $\Gamma^{\alpha}_{\mu\beta}$ by $(w_{\mu})^{a}_{\ b}=e^{a}_{\alpha}\partial_{\mu} e^{\alpha}_b+e^{a}_{\alpha}e^{\beta}_{b}\, \Gamma^{\alpha}_{\mu\beta}$. 

The axial symmetry is maintained at the classical level, or in other words,  the conservation law  $\nabla_\mu j^\mu_A=  0$ holds for any solution of the equations of motion. Quantum mechanically, to check whether the symmetry is maintained one needs to evaluate the  vacuum expectation value of the operator $\nabla_\mu j^\mu_A$. The result, originally computed in \cite{kimura}, is given by 
\be \langle  \nabla_\mu j^\mu_A \rangle =  \frac{2i\hbar}{(4\pi)^2} tr [\gamma_5 E_2(x)] \label{DivE2} \ , \ee
where $E_2(x)$ is the second DeWitt coefficient (see the appendix for a sketch of the derivation, \mbox{and \cite{Parker-Toms}} for a pedagogical calculation using different renormalization methods). In short, the DeWitt coefficients are  local functions constructed from curvature tensors that encode the information of the short distance  behavior ($x' \to x$) of the solution $K(\tau, x, x')$ of a heat-type equation associated with the Dirac operator $D\equiv i\gamma^{\mu}\nabla_{\mu}$ (for~this reason, this function $K$ is called the Heat-Kernel):~
\be i\partial_{\tau} K(\tau, x, x')= D^2 K(\tau, x, x') \ . \label{HeatKerneleq} \ee 

The asymptotic form of $K(\tau, x, x)$ as $\tau \to 0$ defines the $E_n(x)$ coefficients by
\be K(\tau, x, x) \sim \frac{-i}{(4\pi\tau)^2}\sum_{n=0}^{\infty} (i\tau)^n E_n(x) \ . \ee
$E(x)$ are local quantities encoding analytical information of the Klein--Gordon operator $D^2$ in Equation (\ref{HeatKerneleq})
\be D^2\psi= (g^{\mu\nu}\nabla_\mu \nabla_\nu + \mathcal Q(x)) \psi=0 \  \ee
and  are determined by the geometry of the spacetime background.
The  result for the $E_2(x)$  is \cite{Parker-Toms}
\bea \label{E2DW}
E_2(x) & = & \left[-\frac{1}{30} \Box R+\frac{1}{72}R^2-\frac{1}{180}R_{\mu\nu}R^{\mu\nu}+\frac{1}{180}R_{\alpha\beta\mu\nu}R^{\alpha\beta\mu\nu} \right] \mathbb I \label{E2}  \\
 & + & \frac{1}{12}W_{\mu\nu}W^{\mu\nu}+\frac{1}{2}\mathcal Q^2-\frac{1}{6}R \mathcal Q+\frac{1}{6}\Box \mathcal Q
\nonumber \eea
where $W_{\mu\nu}$ is defined by $W_{\mu\nu}\psi=[\nabla_\mu, \nabla_\nu]\psi$, and 
\be\label{QW} \mathcal Q= \frac{1}{4}R \ , \ \ \ \ W_{\mu\nu}= -iR_{\mu\nu \alpha\beta} e^{\alpha}_{a}e^{\beta}_{b}\Sigma^{ab}. \ee

The non-trivial contribution to the axial anomaly  comes entirely  from the $W_{\mu\nu}W^{\mu\nu}$ term and~produces
\bea  \langle  \nabla_\mu j^\mu_A \rangle& =&\frac{2i\hbar}{(4\pi)^2} tr [\gamma_5 E_2(x)]= - \frac{2i\hbar}{(4\pi)^2} \frac{1}{12}R_{\mu\nu ab}R^{\mu\nu cd}tr [\gamma_5 \Sigma^{ab} \Sigma_{cd} ] \nonumber \\
&=&\frac{\hbar}{192\pi^2} R_{\mu \nu \lambda \sigma} {^{\star}R}^{\mu \nu  \lambda \sigma} \ . \eea

If, in addition to the gravitational background, the fermion field propagates also on an electromagnetic background, there is another contribution to the anomaly (this one is proportional to the square of the electric charge $q$ of the fermion). The extra contributions to $W_{\mu\nu} $ and $\mathcal{Q}$ are  $W_{\mu\nu} = iqF_{\mu\nu}$ and $\mathcal Q=2qF_{\mu\nu}\Sigma^{\mu\nu}$, and the expression for $\langle  \nabla_\mu j^\mu_A \rangle $ becomes
\be \langle  \nabla_\mu j^\mu_A \rangle =  \frac{\hbar}{192\pi^2} R_{\mu \nu \lambda \sigma} {^{\star}R}^{\mu \nu  \lambda \sigma} -\frac{\hbar q^2}{8\pi^2}  F_{\mu \nu} {^{\star}}F^{\mu \nu} \ . \ee

To finish this section, recall that there is another type of spin $1/2$ fermions known as  Majorana spinors. They are the ``real''  versions of Dirac's spinors. 
Mathematically, while for Dirac massless fermions the two Weyl spinors $\psi_L$ and $\psi_R$ in Equation (\ref{psi}) are independent of each other, this is not true for Majorana spinors, for which there is an  extra condition $\psi_R= i\sigma^2 \psi_L^*$  \cite{qftbook}. Furthermore, \mbox{the Lagrangian} density for Majorana spinors carries  an additional  normalization factor $1/2$ compared   to Dirac's Lagrangian. Since Majorana spinors do not carry an electric charge ($q=0$), the presence of an electromagnetic background does not induce any anomaly, and the coefficient in the gravitational sector of the anomaly is half of the value obtained for a Dirac fermion.\\

\subsection{Electrodynamics in Curved Spacetime}\label{s.4.2}

Consider Maxwell theory in the absence of electric charges and currents. This theory can be described by a classical action that is formally analog to the action of a Majorana 4-spinor.  Rather than proving from scratch that the familiar Maxwell action can be re-written in the form just mentioned (\mbox{see \cite{ADRNS18}}), we will simply postulate the new action and  show then that it reproduces the correct equations of motion. Consider then the following action in terms of self-dual and anti \mbox{self-dual variables}: 
\be \label{DiracS} S[A^+,A^-]=-\frac{1}{4}\int d^4x\sqrt{-g}\ \bar \Psi \, i\beta^{\mu}\nabla_{\mu}\Psi  \ , \ee
where
\bea \label{Psi}
\Psi=\left( {\begin{array}{c}
 A^+ \\ H_+  \\ A^-  \\ H_- \\
  \end{array} } \right)   \, , \hspace{.5cm} \bar \Psi =   ( A^+,   H_+,    A^-,   H_- ) \, , \hspace{.5cm} \beta^{\mu}= i\,     \left( {\begin{array}{cccc}
 0 & 0 & 0  & \bar\alpha^{\mu}  \\
  0  & 0 &  -\alpha^{\mu}  &0 \\
 0 &  \alpha^{\mu}  & 0  & 0 \\
- \bar  \alpha^{\mu}  & 0  & 0  &0 \\
  \end{array} } \right) 
\ .  \eea 

Note that $\Psi$  is formally analog to  a Majorana 4-spinor rather than a Dirac one, since its lower two components are complex conjugate from the upper ones. Therefore, the action of Equation (\ref{DiracS}) is the analog of Majorana's action.  The independent variables in this action are the potentials $A_{\pm}^{\mu}$, and the fields $\vec H_{\pm}$ are understood as shorthands for their expressions in terms of the potentials (see Section \ref{Electrodynamics}). Note also that Equation (\ref{DiracS}) is a first-order action (i.e.,\ first-order in time derivatives), while the standard Maxwell's action is second order.  $\nabla_{\mu}$ in Equation (\ref{DiracS}) is the covariant derivative acting on the field $\Psi$, given by 
\be \nabla_\mu \Psi = (\partial_\mu + i\omega_{\mu ab}M^{ab}) \Psi \ee
and $M^{ab}$ is 
\bea
M^{ab}=\left( {\begin{array}{cccc}
  \Sigma^{ab} & 0 & 0  & 0  \\
  0  & \,  ^+\Sigma^{ab} & 0  &0 \\
 0 & 0  &  \Sigma^{ab}  & 0 \\
0  & 0  & 0  &  \, ^-\Sigma^{ab} \\
  \end{array} } \right)
\  , \eea 
where $ \Sigma^{\sigma\rho}_{\ \  \alpha\beta}=\delta^{\rho}_{\alpha}\delta^{\sigma}_{\beta} - \delta^{\rho}_{\beta}\delta^{\sigma}_{\alpha}$ is the generator of the $(1/2,1/2)$  representation of the Lorentz group, while $^+\Sigma^{\sigma\rho}_{IJ}$ and  $^-\Sigma^{\sigma\rho}_{\dot I \dot J}$ are the generators of the $(0,1)\oplus (0,0)$ and $(1,0)\oplus (0,0)$  representations, respectively.

Using some  algebraic properties of the  matrices $\alpha$ (see \cite{ADRNS18} for more details), it is not difficult to find that $\beta^{\mu}$ satisfies the  Clifford algebra 
\be \label{Cliff}\{\beta^{\mu},\beta^{\nu} \}=2g^{\mu\nu}\mathbb I \, . \ee
It can also be checked that $\nabla_{\nu}\beta^{\mu}(x)=0$. These matrices can then be thought of as the spin $1$ counterpart of the Dirac $\gamma^{\mu}$ matrices. Furthermore,  one can also introduce   the ``chiral'' $\beta_5$ matrix in a similar way:
\bea \label{beta52}
\beta_5\equiv \frac{i}{4!}\epsilon_{\alpha\beta\gamma\delta}\beta^{\alpha}\beta^{\beta}\beta^{\gamma}\beta^{\delta} =  \left( {\begin{array}{cccc}
- \mathbb I &0  & 0  &0 \\
  0  & -\mathbb I & 0  & 0 \\
 0 & 0 & \mathbb I  & 0 \\
 0 & 0  &  0 & \mathbb I \\
  \end{array} } \right)
\ , \eea
satisfying   properties analogous to the Dirac case:
\bea
\{\beta^{\mu},\beta_5\}=0\, , \hspace{1cm} \beta_5^2=\mathbb I\,  
\ . \eea
Further details and properties  of these matrices can be studied in  \cite{ADRNS18}.

Although the  basic variables in the  action are the potentials $A^{\pm}_{\mu}$, at the practical level one can work by considering $ \Psi$ and $\bar \Psi$ as independent fields. Note that this is the same as one does when working with Majorana spinors. The equations of motion take the form
\be  \label{eqsPsi} \frac{\delta S}{\delta \bar \Psi }=0\hspace{0.5cm} \longrightarrow \hspace{0.5cm}i \beta^{\mu} \nabla_{\mu} \Psi=0 \, .\ee
They contain four equations, one for each of  the four components of $\Psi$. The upper two are the equations  $\bar \alpha^{\mu\nu}_{\hspace{0.15cm}\dot I}\  \nabla_{\mu} A^+_\nu=0$ and $\alpha^{\mu\nu}_{I} \nabla_{\mu} H^I_+=0$.  The lower two are complex conjugated equations. Since these equations are precisely Maxwell's equations written in self- and anti-self-dual variables, \mbox{this proves} that the action of Equation (\ref{DiracS}) describes the correct theory.

Now we study  how the classical electric-magnetic {symmetry} and its related conservation law arise in this formalism. By means of the chiral matrix $\beta_5$, an electric-magnetic duality rotation can be written in the following form,   manifestly analog to a chiral transformation for Dirac fields:
\bea \label{chiraltransf}
\Psi \rightarrow e^{ {}  i\theta \beta_5}\Psi \, , \hspace{1cm} \bar\Psi \rightarrow \bar\Psi e^{ {} i \theta \beta_5} \,. 
\eea

Recalling the explicit form of $\beta_5$ in Equation (\ref{beta52}), one infers that the upper two components of $\Psi$, namely $(A_+, H_+)$, encode the self-dual, or positive chirality sector of the theory, while the lower two components $(A_-, H_-)$ describe the anti-self-dual or the negative chiral sector. The Lagrangian density in Equation (\ref{DiracS}) remains manifestly invariant under these rotations, and in the language of $\Psi$ the Noether current reads
\bea\label{jDdirac}
j_D^{\mu}= \frac{1}{4} \bar \Psi \beta^{\mu}\beta_5 \Psi \ . 
\eea
The corresponding Noether charge  yields
\be
Q_D= \int_{\Sigma_t} d\Sigma_\mu \,  j_D^\mu
= \frac{1}{4} \int_{\Sigma_t}  d\Sigma_3  \, \bar \Psi \beta^{0}\beta_5 \Psi \,  , \ee
where  $d\Sigma_3$ is the volume element of a space-like Cauchy hypersurface $\Sigma_t$. This formula for   $Q_D$ is in full agreement to that  calculated in previous sections (see Equation\ (\ref{Cargadual2})), generalized to curved spacetimes.

The calculation of the vacuum expectation value $ \langle \nabla_{\mu}j_D^{\mu} \rangle$  in the quantum theory  follows exactly the same steps shown above for fermions. 
Namely,   $ \langle \nabla_{\mu}j_D^{\mu} \rangle$ is given again \cite{ADRNS17, ADRNS18} in terms of the second DeWitt coefficient $E_2(x)$
by 
\bea
 \langle \nabla_{\mu}j_D^{\mu} \rangle_=-i\frac{\hbar}{32\pi^2} \, {tr}[\beta_5E_2] \, ,\label{calculation}
\eea
where   $E_2(x)$ is now obtained from the heat kernel $K$ associated with the {\em Maxwell operator} $D=i\beta^{\mu}\nabla_{\mu}$, rather than the Dirac operator $i\gamma^{\mu}\nabla_{\mu}$. The DeWitt coefficient is still given by Equation~(\ref{E2}), but now Equation~(\ref{QW}) needs to be replaced by 

\be \label{Q} \mathcal{Q}\, \Psi\equiv  \frac{1}{2}\beta^{[\alpha}\,  \beta^{\mu]}\, W_{\alpha\mu} \, \Psi\,  \ee
and 
\be  \label{W} W_{\alpha\mu} \Psi \equiv [\nabla_{\alpha},\nabla_{\mu}]\Psi = \frac{1}{2}\, R_{\alpha\mu\sigma\rho}\,  M^{\sigma\rho}\Psi \,  .\ee
With this, Equation (\ref{calculation}) becomes
\be \label{anomaly} \langle \nabla_{\mu} j_D^\mu\rangle_{\rm ren}=-\frac{\hbar}{96\pi^2}\, R_{\alpha\beta\mu\nu} \, ^\star R^{\alpha\beta\mu\nu} \, . \ee

This result reveals that quantum fluctuations spoil the conservation of the axial current $j_D^\mu$  and break the classical symmetry under electric-magnetic (or chiral) transformations, if the spacetime curvature is such that  the curvature invariant $R_{\alpha\beta\mu\nu} \, ^\star R^{\alpha\beta\mu\nu}$ is different from zero.

\section{Discussion}

The  result shown in Equation (\ref{anomaly}) implies that the classical Noether charge $Q_D$ is not necessarily  conserved in the quantum theory, and its change between two instants $t_1$ and $t_2$ can be written as 
\bea
\Delta \left<Q_D\right>= -\frac{\hbar}{96\pi^2}\, \int_{t_{1}}^{t_{2}}\int_{\Sigma} d^4 x \sqrt{-g} \, R_{\alpha\beta\mu\nu} \, ^\star R^{\alpha\beta\mu\nu} = -\frac{\hbar}{6 \pi^2} \int_{t_{1}}^{t_{2}}dt\int_{\Sigma} d^3 x \sqrt{-g} \, E_{\mu\nu}B^{\mu\nu}  \, ,  \label{qanomaly}
\eea
where in the last equality we have written $R_{\alpha\beta\mu\nu} \, ^\star R^{\alpha\beta\mu\nu}$ in terms of the  electric $E_{\mu\nu}$ and magnetic $B_{\mu\nu}$ parts of the Weyl curvature tensor. Note the close analogy with the chiral spin 1/2 anomaly shown in Equation (\ref{DeltaNRNL2}). This result  implies that the polarization state of the quantum electromagnetic field can change in time, even in the complete absence of electromagnetic sources, due to the influence of  gravitational  dynamics and quantum electromagnetic effects (notice the presence of $\hbar$). In this precise sense, one can think about the spacetime as an optically active medium.

  Since  $\Delta \left<Q_D\right>$ is proportional to $\hbar$, one could expect the net effect of the anomaly to be small. \mbox{However, recall} that $ \Delta \left<Q_D\right>= \hbar (N_R - N_L) $. Thus, the net 
  number $N_R-N_L$ is only 
    given by the (dimensionless) geometric integral on the RHS of Equation (\ref{qanomaly}).  
  A sufficiently strong gravitational background could lead  to a significant effect.   
It is also important to remark that Expression (\ref{qanomaly}) accounts for the net helicity created out of an initial {\it vacuum} state---we call this {\em spontaneous} creation of helicity. However, it is well-known in the study of particle creation by gravitational fields that the spontaneous creation effect for bosons always comes together with the {\em stimulated} counterpart, if the initial state is not the vacuum but rather contains quanta on it (see \cite{parkerthesis,agulloparker1,agulloparker2}). The stimulated effect is enhanced by the number of initial quanta. For the same reason, the value of $\Delta \left<Q_D\right>$ is expected to be enhanced if the initial state of radiation is not the vacuum but rather an excited state, as for instance a coherent state which describes accurately the radiation emitted by, say, an astrophysical object. However, remember that the average number of photons in such a coherent state is macroscopic, so it can lead to detectable effects. Therefore, it is conceivable that the change in the polarization of electromagnetic radiation crossing a region of strong gravitational field, produced for instance by the merger of two compact objets, takes macroscopic values. The computation of the exact value of the RHS of Equation (\ref{qanomaly}) in such a situation requires the use of numerical relativity techniques, and this will be the focus of a future {project}.

Finally, we want to mention that the experimental investigation of this anomaly  could be relevant in other areas of physics, as in condensed matter physics \cite{semimetals},  non-linear optics \cite{Leonhardt}, or analogue gravity in general. For instance, metamaterials can be designed to manifest properties that are difficult to find in nature \cite{Pendryetal}. In this case, the medium, and not a distribution of mass-energy, can originate effective geometries \cite{Leonhardt}. They thus may mimic a curved spacetime with optimal values of Equation (\ref{qanomaly})  \mbox{and  could} serve to test the  photon right--left asymmetry originating from the electric-magnetic \mbox{quantum anomaly.}

\vspace{6pt}

\appendix 


\section{Some Details Regarding the Calculation of $\nabla_{\mu}j_A^\mu$ in Curved Spacetimes}
In this appendix, we give a sketch of the derivation of Equation (\ref{DivE2}). The operator  of interest, $\nabla_{\mu}j_A^\mu$, is  quadratic in the fermion fields, and thus  suffers from ultraviolet (UV) divergences. As a consequence, its vacuum expectation value must incorporate renormalization counterterms in order to cancel out  all of them and to provide a finite physically reasonable result:
\be \langle \nabla_{\mu} j_A^\mu\rangle_{\rm ren} =\langle \nabla_{\mu} j_A^\mu\rangle-\langle \nabla_{\mu} j_A^\mu\rangle_{\rm Ad(4)} \, .\ee

Here, $\langle \nabla_{\mu} j_A^\mu\rangle_{\rm Ad(4)} $ denotes the  (DeWitt--Schwinger) asymptotic  expansion up to  the fourth  adiabatic \mbox{order \cite{Parker-Toms}}. Namely, the renormalization method works by expressing $\langle \nabla_{\mu} j_A^\mu\rangle$ in terms of the Feymann two-point function  {$ S(x,x')=-i \langle T \Psi(x) \bar \Psi(x')\rangle$} and then  replacing $S(x,x')$ with $[S(x,x')- S(x,x')_{\rm Ad(4)}]$, where $S(x,x')_{\rm Ad(4)}$ denotes the DeWitt-Schwinger subtractions up to  the fourth  adiabatic order, and finally taking the limit $x\to x'$. 

It is convenient to introduce an auxiliary parameter $s> 0$  in order to regularize spurious infrared divergences in intermediate steps; $s$  will be set to zero at the end of the calculation.  This parameter is introduced by replacing the wave equation  $D\Psi=0$ by  $(D{+}s)  \Psi=0$, where $D\equiv i\gamma^{\mu}\nabla_{\mu}$. As a result,
\bea  \label{jM} \nabla_{\mu} j_A^\mu(x)& =& \nabla_{\mu} \left[ \bar \Psi (x)\gamma^{\mu} \gamma_5 \,\Psi(x)\right] =  -i \left[\bar \Psi (x)\overset{{}_{\leftarrow}}{D}  \, \gamma_5 \,\Psi(x)- \bar \Psi(x) \gamma_5   \overset{{}_{\rightarrow}}{D} \Psi(x) \right] \nonumber \\ &=& \lim_{\substack{s \to 0 \\ x \to x'}} -2i \, s\,  \bar \Psi(x) \gamma_5  \Psi(x')=\lim_{\substack{s \to 0 \\ x \to x'}} -2i \,s\,  {\rm Tr}[ \gamma_5  \Psi (x) \bar \Psi(x') ]\, , \eea
where we have used $\{\gamma^{\mu},\gamma_5\}=0$. Picking up an arbitrary vacuum state  $|0\rangle$, we have
\be \label{formal} \langle \nabla_{\mu} j_A^\mu\rangle= \lim_{\substack{s\to 0 \\ x \to x'}} \, 2 \,  s\,  {\rm Tr}\Big[ \gamma_5 \, S(x,x',s)\Big]\, ,  \ee 
and the renormalized expectation value is given by
\be \label{renj} \langle \nabla_{\mu} j_A^\mu\rangle_{\rm ren}= \lim_{\substack{s\to 0\\ x \to x'}} \, \frac{ 1 }{2} \,  s\,  {\rm Tr}\Big[ \gamma_5 \, \Big(S(x,x',s)-S(x,x',s)_{\rm Ad(4)}\Big)\Big]\, . \ee 
Here, $S(x,x',s)$ encodes the information of the  vacuum state, and  the role of $S(x,x',s)_{\rm Ad(4)}$ is to remove the  ultra-violet divergences---which are the same regardless of the choice of vacuum. It  is now useful to write $S(x,x',s)_{\rm Ad(4)}=   \left[ (D_x-s) G(x,x',s)\right]_{\rm Ad(4)}$, since it is known that \cite{Parker-Toms}
\bea  \label{GAd} G(x,x',s)\sim  \frac{\hbar \Delta^{1/2}(x,x')}{16\pi^2} \sum_{k=0}^{\infty} E_k(x,x')  \int_0^{\infty} d\tau \, e^{-i\, (\tau  s^2+\frac{\sigma(x,x')}{2 \tau})}\, (i\tau)^{(k-2)} \,  \label{asympG}\, .\eea
In this expression, $\sigma(x,x')$ represents   half of the geodesic distance squared between $x$ and $x'$, $\Delta^{1/2}(x,x')$ is the Van Vleck-Morette determinant, and $E_k(x,x')$ are the DeWitt coefficients introduced in the main text ($E_k(x)\equiv \lim_{x'\to x}E_k(x,x')$). 

We can safely take now the limit  $x=x'$ in which the two points merge.  
Due to the symmetry of the classical theory, the bare contribution  $S(x,x',s)$ in Equation (\ref{renj}) vanishes for any choice of vacuum state. As a result, $\langle \nabla_{\mu} j_A^\mu\rangle_{\rm ren}$ arises entirely from the subtraction terms, $S(x,x',s)_{\rm Ad(4)}$.  
This means that   $\langle \nabla_{\mu} j_A^\mu\rangle_{\rm ren}$ is {\em independent of the choice of vacuum}. On the other hand, it is not difficult to see that only the terms with $k=2$ in Equation (\ref{GAd}) produce a non-vanishing result. Additionally,  terms  involving derivatives of $E_2(x,x')$ must be disregarded because they involve five derivatives of the metric and hence   are of  the  fifth  adiabatic order. With all these considerations, Expression (\ref{renj}) leads then to Formula (\ref{DivE2}).

\acknowledgments{ We thank A. Ashtekar,  P. Beltran, E. Bianchi, A. Ferreiro, S. Pla, and J. Pullin for useful~discussions. This research was funded with  Grants. No.\ FIS2014-57387-C3-1-P,  No.\  FIS2017-84440-C2-1-P, No. \ FIS2017-91161-EXP, No. \  SEV-2014-0398, and No. \ SEJI/2017/042 (Generalitat Valenciana),    COST action CA15117 (CANTATA), supported by COST (European Cooperation in Science and Technology)
, NSF CAREER Grant No.\ PHY-1552603, ERC Consolidator Grant 
No. MaGRaTh-64659 }










\end{document}